\documentstyle[preprint,eqsecnum,aps]{revtex}
\tighten
\begin{document}
\draft
\preprint{DTP/94/08}
\title{QCD predictions for the transverse energy flow \\  in deep-inelastic
scattering in the HERA small $x$ regime
}
\author{J.\ Kwieci\'{n}ski \footnotemark[1], A.D.\ Martin \footnotemark[2],
P.J.\ Sutton \footnotemark[3] and K.\ Golec-Biernat \footnotemark[1]
\vspace*{2mm}
}
\address{
\footnotemark[1] Department of Theoretical Physics, H.\ Niewodniczanski
Institute of Nuclear Physics,\\ ul.\ Radzikowskiego 152, 31-342
Krakow, Poland. \\ \vspace*{2mm}
\footnotemark[2]
Department of Physics, University of Durham, DH1 3LE, England.
\\ \vspace*{2mm}
\footnotemark[3]
Department of Physics, University of Manchester, M13 9PL, England.
\vspace*{5mm}
}
\date{11th March 1994}
\maketitle
\begin{abstract}
The distribution of transverse energy, $E_T$, which accompanies deep-inelastic
electron-proton scattering at small $x$, is predicted in the central region
away  from the current jet and proton remnants.  We use BFKL dynamics, which
arises  from the summation of multiple gluon emissions at small $x$, to derive
an  analytic expression for the $E_T$ flow.  One interesting feature is an
$x^{-\epsilon}$ increase of the $E_T$ distribution with decreasing $x$, where
$\epsilon = (3\alpha_s/\pi)2\log 2$.  We perform a numerical study to
examine the possibility of using characteristics of the $E_T$ distribution as a
means of identifying BFKL dynamics at HERA.
\end{abstract}
\section{Introduction}
\label{sec:one}

The experiments at HERA measure deep-inelastic electron-proton
scattering in the previously unexplored small $x$ regime, $x
\lesssim 10^{-3}$.  As usual, $x$ is the Bjorken variable
$x=Q^2/2p\cdot q$ and $Q^2=-q^2$, where $p$ and $q$ are the four
momenta of the incoming proton and the virtual photon probe
respectively. The deep-inelastic observables reflect the small
$x$ behaviour of the gluon, which is by far the dominant parton
in this kinematic region.  In particular the small $x$ behaviour
of the structure function $F_2(x,Q^2)$ is driven by the $g\to
q\bar{q}$ transition.  At such values of $x$ soft gluon emission
and the associated virtual gluon corrections give rise to powers
of  $\log (1/x)$, which have to be resummed.  To leading
$\alpha_s \log (1/x)$ order, this is carried out by the BFKL (or
Lipatov) equation\cite{BFKL} which, in a physical gauge,
effectively amounts to the summation of gluon ladder diagrams.
Two characteristic features of the result of this procedure are,
first, the $x^{-\lambda}$ growth of the gluon distribution
$xg(x,Q^2)$ as $x$ decreases, with $\lambda \sim  0.5$, and,
second, the relaxation of the strong-ordering of the transverse
momenta, $k_T$, of the gluons along the ladder (which is
characteristic of the ``large'' $x$, large $Q^2$
``Altarelli-Parisi'' regime).

   The recent measurements of $F_2(x,Q^2)$ at HERA\cite{H1,ZEUS}
show a rise with decreasing $x$ which is entirely consistent with
the $x^{-\lambda}$ BFKL perturbative QCD prediction\cite{AKMS}.  However
the observed small $x$ behaviour of $F_2$ can also be
mimicked by conventional dynamics based on the Altarelli-Parisi
equation\cite{AGKMS}. Because of the inclusive nature of $F_2$,
it is not a sensitive discriminator between BFKL
and conventional dynamics.  For this purpose it is necessary
to look into properties of the final state.  Here we investigate
whether or not the emitted transverse energy, $E_T$, which
accompanies a deep-inelastic event (see Fig. 1), can be used to identify
the BFKL dynamics.  Due to the relaxation of the strong-ordering
of the gluon $k_T$'s along the chain we expect to find more transverse energy
in the central region (between the current jet and the proton  remnants) than
would result from conventional evolution.  Indeed a Monte Carlo study, in the
context of heavy quark  production, of the $E_T$ distribution as a function of
rapidity hints at an  increase of the $E_T$ flow if QCD \lq\lq all-loop"
dynamics  \cite{CIAF,MAR1,MAR2} (which incorporates BFKL effects) is employed,
see Fig.\ 9  of Ref.\ \cite{BRW}.

Now the $E_T$ emitted by the gluon chain is given by the integral over the
inclusive distribution for the emission of one gluon weighted by its $E_T$.
Thus, for a given element of $(x,Q^2)$, the distribution of transverse energy
as a function of $\log x_j$ is given by
\begin{equation}
x_j \frac{\partial E_T}{\partial x_j}\; = \; {1\over \sigma}\int dk{_j^2} x_j
 \frac{\partial \sigma}
{\partial x_j\partial k_j^2} |{\bf k_j}|
\label{aa}
\end{equation}
where $\sigma $ denotes the cross section in the
$(x,Q^2)$ interval, and where $x_j$ and $k_j$ are the
longitudinal momentum fraction and the transverse
momentum carried by the gluon jet in the photon-proton
centre-of-mass frame.  The variable $\log x_j$ is closely related to rapidity,
see Sec.\ \ref{sec:twob}.  In terms of the proton structure functions,
the differential cross section in Eq.\ (\ref{aa}) is
\begin{equation}
\frac{\partial \sigma}{\partial {x_j}\partial k_j^2}
= \frac{4\pi {\alpha ^2}}
{xQ^4}\left[ (1-y)\frac{\partial F_2}{\partial x_j
\partial k_j^2} + \frac{1}{2}y^2\frac {\partial (2xF_1)}
{\partial x_j\partial k_j^2}\right]
\label{ab}
\end{equation}
where $y=Q^2/xs$ with $\sqrt s$ being the centre-of-mass
energy of the colliding $ep$ system.

  With these definitions, we can calculate the energy
flow, $x_j \partial E_T/\partial x_j$ of Eq.\ (\ref{aa}), by using
the diagrams of Fig. 2 to evaluate the differential
structure functions.  We have \cite{GLR,LEV1,LEV2}
\begin{equation}
x_j\frac{\partial F_i}{\partial x_j\partial k^2_j} =
\int \frac{d^2k_p}{\pi k^4_p}\int \frac {d^2k_{\gamma}}
{k^4_{\gamma}} \left(\frac {3\alpha _s}{\pi}
\frac {k^2_p k^2_{\gamma}}{k^2_j}\right)
{\cal F}_i(x/x_j,k^2_{\gamma},Q^2) f(x_j,k^2_p)
\delta ^{(2)}(k_j-k_{\gamma}-k_p)
\label{ac}
\end{equation}
with $i=1,2$, where ${\cal F}_i$ and $f$ describe,
respectively the soft gluon resummation above and
below the emitted gluon of Fig. 2.  That is ${\cal F}_i$
and $f$ satisfy BFKL equations, provided of course that
$x/x_j$ and $x_j$ are sufficiently small.  The BFKL equation for $f$ may be
written in the differential form
\begin{equation}
-z \frac{\partial f(z,k^2)}{\partial z} = \frac{3\alpha_s}{\pi} k^2 \int
\frac{dk^{\prime 2}}{k^{\prime 2}} \left[ \frac{f(z,k^{\prime 2}) -
f(z,k^2)}{|k^{\prime 2} - k^2|} + \frac{f(z,k^2)}{(4k^{\prime 4} +
k^4)^{\frac{1}{2}}} \right] \equiv K \otimes f,
\label{ad}
\end{equation}
with a similar form \cite{KMS} for ${\cal F}_i$.   As usual,
$f$ is the unintegrated gluon distribution in the proton,
that is with the final integration over $dk^2_p$ not performed. The expression
in brackets in (\ref{ac}) arises from the (square of the) BFKL
vertex for real gluon emission, where $k_p, k_{\gamma}$
and $k_j$ are the gluon transverse momenta in the
photon-proton frame.

We may use Eqs (\ref{aa})--(\ref{ac}), together with the
solutions of the BFKL equations  for $f$  and ${\cal F}_i$, to
determine the $E_T$ flow,  $\partial E_T/\partial \log x_j$,
which accompanies deep-inelastic scattering, provided $x \ll x_j
\ll  1$.  Valuable insight can be obtained by using the analytic
solutions of the  BFKL equation for fixed $\alpha_s$.  Therefore,
before presenting the full  numerical calculation, we perform in
Sec.\ \ref{sec:two}  an analytic study to establish  general
characteristics of the $E_T$ distribution.  First, we make the
simplifying assumption of strong-ordering at the vertex for real
gluon emission;  in fact in addition to the \lq\lq normal"
strong-ordering, $k^2_j \gg k^2_p$, in  the expression in
brackets in Eq.\ (\ref{ac}),  we also need to include the
contribution from  the \lq\lq anomalous" ordering $k^2_p \gg
k^2_j$.   Then we present an exact  analytic study by using the
BFKL form of the vertex, which incorporates all  orderings
automatically.   In Sec. \ref{sec:four}  we present the
predictions of the $E_T$  distribution obtained when the BFKL
equation is solved numerically.  At each  stage we compare with
the $E_T$ flow pattern predicted by coventional dynamics  based
on the Altarelli-Parisi equation. Section \ref{sec:five} contains
our conclusions.

\section{Characteristics of $E_T$ flow: an analytic study}
\label{sec:two}

At first sight the prediction for the shape of the $E_T$ distribution as a
function of $\log x_j$ looks simple.  The leading small $z$ behaviour of the
solutions of the BFKL equation for fixed $\alpha_s$ have the form
\begin{equation}
f(z,k^2) \sim z^{-\lambda}, \hspace*{1cm} \text{with} \hspace*{.6cm} \lambda =
4\bar{\alpha}_s \log 2
\label{ba}
\end{equation}
where $\bar{\alpha}_s \equiv 3\alpha_s/\pi$, and similarly for ${\cal F}_i$.
Thus for  $x/x_j$ and $x_j$ sufficiently
small, we see that the relevant combination which occurs in
Eq.\ (\ref{ac}) has the behaviour
\begin{equation}
\frac {1}{\sigma}{\cal F}_i(x/x_j,...)f(x_j,..)\sim
 \frac {1}{x^{-\lambda}}\left(\frac{x}{x_j}\right)^{-\lambda}
(x_j)^{-\lambda}
\label{bb}
\end{equation}
which appears to give an $E_T$ distribution which is independent of $x_j$, and
also of  $x$.  Thus well away from
both the current jet and the proton remnants, we
anticipate a plateau distribution of $E_T$ as a function
of $ \log x_j$, which remains unchanged as $x$ varies.

However the situation is more subtle.   We have overlooked the $x$ and $x_j$
dependences which arise from the integrations over the transverse momenta.  To
see this we need to look more closely at the analytic solution of the BFKL
equation.  Recall that the solution may be obtained by taking the Mellin
transform $\tilde{f}$ of $f$:
\begin{equation}
f(z,k^2) \; = \; \frac{1}{2\pi i} \int^{c + i\infty}_{c - i\infty}
(k^2)^{\omega} \tilde{f}(z,\omega) d\omega
\label{bc}
\end{equation}

\begin{equation}
\tilde{f}(z,\omega) \; = \; \int^{\infty}_0 (k^2)^{-\omega - 1} f(z,k^2) dk^2
\label{bd}
\end{equation}
with $c = \frac{1}{2}$.  If we substitute Eq.\ (\ref{bd}) into (\ref{ad})
we obtain
\begin{equation}
-z \frac{\partial \tilde{f}(z,\omega)}{\partial x} \; = \; \tilde{K}(\omega)
\tilde{f}(z,\omega)
\label{be}
\end{equation}
where $\tilde{K}$, the transform of the BFKL kernel, is
\begin{equation}
\tilde{K}(\omega) \; = \; \bar{\alpha}_s [2\Psi(1) - \Psi(\omega) -
\Psi(1-\omega)]
\label{bf}
\end{equation}
with $\Psi(\omega) \equiv \Gamma^{\prime}(\omega)/\Gamma(\omega)$.  On
substituting the solution of Eq.\ (\ref{be}) into Eq.\ (\ref{bc}) we have
\begin{equation}
f(z,k^2) \; = \; \frac{1}{2\pi i} \int^{c + i\infty}_{c - i\infty} d\omega
(k^2)^{\omega} \tilde{f}(1,\omega) z^{-\tilde{K}(\omega)} .
\label{bg}
\end{equation}
As $\omega$ varies along the contour of integration, the maximum value of
$\tilde{K}(\omega)$ is found to occur at  $\omega = \frac{1}{2}$.  The region
$\omega \sim \frac{1}{2}$ therefore dominates the small $z$ behaviour of $f$.
To obtain the leading behaviour we expand the various terms in
Eq.\ (\ref{bg}) about the point $\omega = \frac{1}{2}$ and find
\begin{equation}
f(z,k^2) \approx z^{-\lambda} \left( \frac{k^2}{k^2_0} \right)^{\frac{1}{2}}
\frac{\bar{f}_0(\frac{1}{2})}
{[2\pi\lambda^{\prime\prime}
\log (1/z)]^{\frac{1}{2}}} \exp\left(
\frac{-\log ^2(k^2/\bar{k}^2)}
{2\lambda^{\prime\prime}\log (1/z)} \right) ,
\label{bh}
\end{equation}
with an analogous relation for the ${\cal F}_i$
\begin{equation}
{\cal F}_i(z,k^2,Q^2) \approx z^{-\lambda} \left( \frac{k^2}{Q^2}
\right)^{\frac{1}{2}} Q^2 \frac{\bar{{\cal F}}_i^{(0)}(\frac{1}{2})}
{[2\pi\lambda^{\prime\prime}\log (1/z)]^{\frac{1}{2}}} \exp \left(
\frac{-\log ^2(k^2/aQ^2)}{2\lambda^{\prime\prime}\log (1/z)} \right)
\label{bi}
\end{equation}
where $\bar{f}_0(\frac{1}{2}), \bar{{\cal F}}_i^{(0)}(\frac{1}{2}),
k^2_0,\bar{k}^2$ and $a$ are given in terms of the boundary conditions.   The
quantities $\lambda$ and $\lambda^{\prime\prime}$ arise from the expansion of
$\tilde{K}(\omega)$ of Eq.\ (\ref{bf}) about the point $\omega = \frac{1}{2}$:
\begin{equation}
\tilde{K}(\omega) \; = \; \lambda + {\textstyle \frac{1}{2}}
\lambda^{\prime\prime}(\omega - {\textstyle \frac{1}{2}})^2 + ...
\label{bj}
\end{equation}
with $\lambda$ given by Eq.\ (\ref{ba}) and $\lambda^{\prime\prime} =
\bar{\alpha}_s 28\zeta(3)$,
where the Riemann zeta function $\zeta(3) \approx 1.2$.

Note that the scale of the ${\cal F}_i$ is provided by $Q^2$ and also that,
with  our definition \cite{KMS}, the ${\cal F}_i$ have the dimensions of $Q^2$.
In  Eq.\ (\ref{bh}) we have used a slightly different notation for the function
$\tilde{f}(1,\omega)$ at the boundary (which for simplicity we take to be $z =
1$).  We have introduced a dimensionless boundary function $\bar{f}_0$ defined
by
\begin{equation}
\tilde{f}(1,\omega) \; = \; (k^2_0)^{-\omega} \bar{f}_0(\omega) .
\label{bk}
\end{equation}
Similarly for the ${\cal F}_i$ we have introduced  dimensionless boundary
functions $\bar{{\cal F}}_i^{(0)}$ defined by
\begin{equation}
\tilde{{\cal F}}_i(1,\omega,Q^2) \; = \; (Q^2)^{1-\omega} \bar{{\cal
F}}_i^{(0)}(\omega) .
\label{bl}
\end{equation}

To estimate the characteristics of the $E_T$ flow we may omit the longitudinal
structure function and assume $2xF_1 = F_2$.  It is straightforward to include
the small correction arising from $F_L = 2xF_1 - F_2$.  Thus (\ref{aa}) becomes
\begin{equation}
x_j \frac{\partial E_T}{\partial x_j} \; = \; \frac{1}{F_2} \int dk^2_j x_j
\frac{\partial F_2}{\partial x_j\partial k^2_j} |{\bf k_j}| ,
\label{bm}
\end{equation}
where the integrand is specified by (\ref{ac}) with $f$ and
${\cal F}_2$  given by Eq.\ (\ref{bh})  and Eq.\ (\ref{bi})
respectively,  for sufficiently small $x_j$ and $x/x_j$.  Thus to
determine the $E_T$ flow we must perform the integrations
$dk^2_j, \, d^2k_p, \,  d^2k_{\gamma}$ over the transverse
momenta, subject of course to the  conservation of transverse
momentum as embodied in the delta function in (\ref{ac}).

\subsection{Simplified analytic treatment}
\label{sec:twoa}

We can obtain valuable insight by making the assumption that the transverse
momenta are strongly-ordered at the vertex for real gluon emission.  At first
sight, we would expect that this meant just keeping the contribution with
$k^2_j  \gg k^2_p$ in Fig.\ 2.  However for the $E_T$ weighted distribution we
shall  find that there is an equally big contribution coming from the region
with  \lq\lq anomalous" ordering $k^2_p \gg k^2_j$.  We evaluate these two
contributions in turn.

For the \lq\lq normal" strong-ordered case with $k^2_j \gg k^2_p$
we have  $k^2_{\gamma} \approx k^2_j$; then  Eq.\ (\ref{ac})
simplifies, since the $d^2k_p  /k^2_p$ integration over $f$ just
gives the integrated gluon distribution $x_j  g(x_j,k^2_j)$.  In
fact in this approximation we recover the formula for
deep-inelastic + energetic jet events
\begin{equation}
x_j\frac{\partial F_2}{\partial x_j\partial k^2_j} =
\left(\frac {\bar{\alpha}_s}{k^4_j}\right) x_j g(x_j,k^2_j)
{\cal F}_2(x/x_j,k^2_j,Q^2),
\label{bn}
\end{equation}
see, for example,
Eq.\ (\ref{bd}) of Ref.\ \cite{KMS}. Using the explicit solutions (\ref{bh})
and (\ref{bi}), and
inserting (\ref{bn}) into (\ref{bm}), we obtain
\begin{equation}
x_j \frac{\partial E_T^{(a)}}{\partial x_j} \; = \; \frac{2\bar{\alpha}_s}{F_2}
x_j^{-\lambda} \left(\frac{x}{x_j}\right)^{-\lambda} \left(
\frac{Q^2\bar{k}^2}{k^2_0} \right)^{\frac{1}{2}} \frac{\bar{f}_0(\frac{1}{2})
\bar{{\cal F}}_2^{(0)}(\frac{1}{2})
I(x,x_j,Q^2)}{[2\pi\lambda^{\prime\prime}\log (1/x_j)2\pi\lambda^{\prime
\prime}\log (x_j/x)]^{\frac{1}{2}}}
\label{bo}
\end{equation}
where
\begin{equation}
I \; = \; \int^{\infty}_0 \frac{dk^2_j}{k^2_j} \frac{|{\bf
k_j}|}{(\bar{k}^2)^{\frac{1}{2}}} \exp \left( -\frac{
\log ^2(k^2_j/\bar{k}^2)}
{2\lambda^{\prime \prime}\log (1/x_j)} -
\frac{\log ^2(k^2_j/aQ^2)}{2\lambda^{\prime\prime}\log (x_j/x)} \right) .
\label{bp}
\end{equation}
The factor 2 in Eq.\ (\ref{bo}) arises when we translate formula (\ref{bh})
for $f(x_j,k^2_p)$
into one for $x_jg(x_j,k^2_j)$.  This factor can be inferred by using Mellin
transform techniques and noting that the main contribution occurs when
$1/\omega  \approx 2$.

The other contribution, which we denote $x_j\partial E_T^{(b)}/\partial x_j$,
coming from the \lq\lq anomalous" ordered region $k^2_p \gg k^2_j$, can be
readily shown to be equal to $x_j\partial E_T^{(a)}/\partial x_j$.  To see
this,  we note that for this case $k^2_{\gamma} \approx k^2_p$ and so we have
\begin{displaymath}
x_j \frac{\partial E_T^{(b)}}{\partial x_j} \; = \; \frac{\bar{\alpha}_s}{F_2}
\int \frac{dk^2_p}{(k^2_p)^2} 2(k^2_p)^{\frac{1}{2}} {\cal
F}_2(x/x_j,k^2_p,Q^2)f(x,k^2_p) ,
\end{displaymath}
where the factor $2(k^2_p)^{\frac{1}{2}}$ arises from
\begin{displaymath}
\int^{k^2_p}_0 (k^2_j)^{-\frac{1}{2}} dk^2_j \; = \; 2(k^2_p)^{\frac{1}{2}} .
\end{displaymath}
If we insert Eqs (\ref{bh}) and (\ref{bi})  then we find
$x_j\partial E_T^{(b)}/\partial x_j$ is  also given by Eq.\
(\ref{bo}).   This so-called anomalous contribution \cite{LVG}
comes from  ordering of transverse momenta that are opposite to
that in the ordinary  Altarelli-Parisi evolution of the structure
function.  We have found that in the  small $x$ limit and for
fixed (albeit large) $Q^2$, that it gives a contribution  to the
$E_T$ flow equal to that arising from the normal strong-ordering.
Why  does this \lq\lq anomalous" contribution not occur, for
instance, in the  conventional double leading logarithm (DLL)
limit of, say, $F_2(x,Q^2)$?  In the  total cross section (or in
an inclusive structure function) the dominant or  leading \lq\lq
anomalous" contribution is cancelled by the virtual corrections
which lead to gluon reggeisation \cite{BFKL,LEV1} (or,
equivalently, to the  \lq\lq non-Sudakov form factor" of Ref.\
\cite{CIAF,MAR1,MAR2}).  The  cancellation ensures that, at small
$x$, we do not encounter the double  logarithmic terms of the
form $\alpha_s  \log ^2(1/x)$.  Moreover the virtual  corrections
guarantee that the conventional DLL terms $\alpha_s \log (1/x)
\log Q^2$ only come from the usual strongly-ordered region of
transverse  momenta, i.e.\ $k^2_p \ll k^2_j \ll Q^2$.  It may
also be interesting to observe  that, unlike the unweighted
distributions, the $E_T$ flow is an infrared finite  quantity
since the factor of $|{\bf k}_j|$ in the integrand of (\ref{aa})
removes the  unintegrable singularity which would otherwise occur
at $k^2_j = 0$, see (\ref{ac}).   After this aside, we now return
to complete the analytic calculation of the  $E_T$ flow.

The integral $I$ of Eq.\ (\ref{bp}) can  readily be evaluated if we
change the variable from
$k^2_j$ to $t =$ $ \log (k^2_j/\bar{k}^2)$.  Then Eq.\ (\ref{bp}) becomes
\begin{equation}
I \; = \; \int^{\infty}_{-\infty} dt \, \exp \left( \frac{t}{2} - \frac
{t^2}{2\lambda^{\prime\prime}\log (1/x_j)} -
\frac{(t-T)^2}{2\lambda^{\prime\prime}\log (x_j/x)} \right)
\label{bq}
\end{equation}
where $T \equiv \log (aQ^2/\bar{k}^2)$.  If we evaluate this
Gaussian  integral, and insert the result into Eq.\ (\ref{bo})
and into the identical expression for  $x_j \, \partial
E_T^{(b)}/\partial x_j$, we finally obtain the following
analytic prediction for the $E_T$ flow
\begin{equation}
x_j \frac{\partial E_T}{\partial x_j} \; = \; x_j \frac{\partial
E_T^{(a)}}{\partial x_j} + x_j
\frac{\partial E_T^{(b)}}{\partial x_j} \; = \; 4
\bar{\alpha}_s (Q^2\bar{k}^2)^{\frac{1}{4}} x^{-\epsilon} \exp \left( -
\frac{\epsilon [\log (x^2_j/x) + 4T/\lambda^{\prime\prime}]^2}
{\log (1/x)} \right) ,
\label{br}
\end{equation}
provided $x/x_j$ and $x_j$ are sufficiently small, where $\epsilon =
\lambda^{\prime\prime}/32$.

To obtain Eq.\ (\ref{br}) we have used the BFKL prediction for $F_2$
in the leading $\log 1/x$ approximation \cite{AKMS}
\begin{equation}
F_2(x,Q^2) \; = \; \int \frac{dk^2}{k^4} f(x,k^2) {\cal F}^{(0)}_2 (k^2,Q^2)
\label{bs}
\end{equation}
where ${\cal F}_2^{(0)}$ is the quark box (and crossed box) contribution shown
in Fig.\ 2(b), and where the integration over the gluon longitudinal momentum
has been performed.  If we insert expression (\ref{bh}) for $f$ and
we take the small
$x$ limit (where the Gaussian factor may be neglected), then the integral
becomes proportional to the Mellin transform $\tilde{{\cal F}}_2^{(0)}$ at
$\omega = \frac{1}{2}$.  Using Eq.\ (\ref{bl}) we obtain
\begin{equation}
F_2 \; = \; x^{-\lambda} \left( \frac{Q^2}{k^2_0}\right)^{\frac{1}{2}}
\frac{\bar{f}_0(\frac{1}{2}) \bar{{\cal
F}}_2^{(0)}(\frac{1}{2})}{[2\pi\lambda^{\prime\prime}
\log (1/x)]^{\frac{1}{2}}} .
\label{bt}
\end{equation}
This result for $F_2$ has been inserted into Eq.\ (\ref{bo}).

Equation (\ref{br})
for the $E_T$ flow applies in the central region away from the
current  jet and the proton remnants, that is when $x_j/x \gg 1$ and $1/x_j \gg
1$.   Contrary to our initial expectation, (\ref{bb}),
the expression (\ref{br}) depends on
$x$,  and also on $x_j$.  The main characteristics of the $E_T$ flow can be
readily  identified from Eq.\ (\ref{br}).
We see that the $E_T$ distribution has a broad
Gaussian  shape in $ \log x_j$ (or in rapidity, see Sec.\ \ref{sec:twob}
below) with
\begin{itemize}
\item[(i)] a peak which moves away from $x_j \sim x^{\frac{1}{2}}$ as $Q^2$
increases,
\item[(ii)] a width which grows as $\sqrt{\log (1/x)}$ as $x$ decreases,
\item[(iii)] an $x^{-\epsilon}$ dependence such that it grows with decreasing
$x$,
\item[(iv)] a $(Q^2)^{\frac{1}{4}}$ dependence such that it grows with
increasing $Q^2$.
\end{itemize}
Although the shape of the distribution is specified, the
normalisation is  controlled by   $ \log (\bar{k}^2)$, the
parameter which gives the centre of the  input (non-perturbative)
$ \log k^2$ Gaussian distribution of $f$,  see Eq.\ (\ref{bh})
and Ref.\ \cite{AKMS}.

\subsection{$E_T$ flow as a function of rapidity}
\label{sec:twob}

We have calculated the $E_T$ flow as a function of $ \log x_j$, a
variable which  is closely related to the rapidity, $y$.  The
result, $\partial E_T/\partial  \log  x_j$, is the Gaussian form
of Eq.\ (\ref{br}).  For practical purposes the  rapidity
distribution $\partial E_T/\partial y$ is more useful, so here we
show  how to evaluate the integral $I$ of Eq.\ (\ref{bq}) in
terms of $y$, rather than  $ \log x_j$.  In this way we will find
that the $E_T$ flow as a function of rapidity also has  a
Gaussian form.

We define the rapidity $y$ in the photon-proton c.m.\ frame in the standard way
\begin{equation}
y \; = \; {\textstyle \frac{1}{2}} \log \left( \frac{x^2_j}{x}
\frac{Q^2}{k^2_j} \right) ,
\label{bu}
\end{equation}
which leads to the following relation between $y$ and $ \log x_j$
\begin{equation}
\log x_j \; \simeq \; y - {\textstyle \frac{1}{2}} (T-t) - {\textstyle
\frac{1}{2}} \log (1/x) .
\label{bv}
\end{equation}
In terms of $y$, the integral $I$ of Eq.\ (\ref{bq}) assumes the
(non-Gaussian) form
\begin{equation}
I \; = \; \int^{t_2}_{t_1} dt \, \exp \left( \frac{t}{2} -
\frac{t^2}{\lambda^{\prime\prime}[-2y + T-t + \log (1/x)]} -
\frac{(t-T)^2}{\lambda^{\prime\prime}[2y - T+t + \log (1/x)]} \right)
\label{bw}
\end{equation}
where the limits
\begin{equation}
t_{1,2} \; = \; T - 2y \mp \log (1/x) .
\label{bx}
\end{equation}
Now, to obtain an analytic form we approximate the $t$ factors in the
denominators in Eq.\ (\ref{bw}) by
\begin{displaymath}
\bar{t} \; = \; 4 \epsilon \, \log (1/x) + {\textstyle \frac{1}{2}} T,
\end{displaymath}
which denotes the value of $t$ for which the integrand in Eq.\ (\ref{bq})
is a maximum in
the leading $ \log (1/x)$ approximation.  It is straightforward to evaluate the
integral approximated in this way.  We find
\begin{equation}
\frac{\partial E_T}{\partial y} \; = \;
2\bar{\alpha}_s(Q^2\bar{k}^2)^{\frac{1}{4}} x^{-\epsilon} \exp \left(
\frac{-\epsilon [2y + 4T/\lambda^{\prime\prime} - \frac{1}{2}T + \frac{1}{8}
\lambda^{\prime\prime}\log (1/x)]^2}{\log (1/x)} \right) .
\label{by}
\end{equation}
Thus we obtain a Gaussian $E_T$ distribution as a function of rapidity $y$,
identical in structure to that in terms of $ \log x_j$,  except for the two
non-leading terms in the exponential.

\subsection{Improved analytic treatment}
\label{sec:twoc}

Equation (\ref{br}), though valuable to see the origin of the $x$
and $x_j$ dependence  of $E_T$, is based on assumptions which are
not necessary even for an analytic  treatment.  In this
subsection we show that it is not necessary to assume (i)
strong-ordering at the gluon emission vertex and (ii) the
approximate Gaussian  diffusion patterns of Eq.\ (\ref{bh}) and
Eq.\ (\ref{bi}).   The procedure is to evaluate
(\ref{aa})--(\ref{ac}) using  Mellin transform techniques.  We
have
\begin{displaymath}
x_j \frac{\partial E_T}{\partial x_j} \; = \; \frac{1}{F_2} \int
\frac{dk^2_p}{k^2_p} \int \frac{dk^2_{\gamma}}{k^2_{\gamma}} \int
\frac{d\phi}{2\pi} \frac{\bar{\alpha}_s}{(k^2_p + k^2_{\gamma} +
2k_pk_{\gamma}\cos \phi)^{\frac{1}{2}}} {\cal F}_2(x/x_j, k^2_{\gamma},Q^2)
f(x_j,k^2_p)
\end{displaymath}

\begin{equation}
\equiv \; \frac{1}{F_2} \int \frac{dk^2_p}{k^2_p} \int
\frac{dk^2_{\gamma}}{k^2_{\gamma}} \frac{1}{(k^2_pk^2_{\gamma})^{\frac{1}{4}}}
H(k^2_{\gamma}/k^2_p) {\cal F}_2(x/x_j,k^2_{\gamma},Q^2) f(x_j,k^2_p) .
\label{bz}
\end{equation}
The function $H$, introduced as above, is now replaced by the inverse Mellin
transform
\begin{equation}
H(k^2_{\gamma}/k^2_p) \; = \; \frac{1}{2\pi i} \int^{c + i\infty}_{c - i\infty}
d\nu \tilde{H}(\nu) \left( \frac{k^2_{\gamma}}{k^2_p} \right)^{\nu}
\label{bza}
\end{equation}
with $c$ in the range $-\frac{1}{4} < c < \frac{1}{4}$.  Proceeding in this way
we see that the $k^2_p$ and $k^2_{\gamma}$ integrations, over $f$ and ${\cal
F}_2$ respectively, now factorize and, moreover, are simply of the form of the
Mellin transforms $\tilde{f}$ and $\tilde{{\cal F}}_2$,
defined as in Eq.\ (\ref{bd}).  Thus Eq.\ (\ref{bz}) becomes
\begin{equation}
x_j \frac{\partial E_T}{\partial x_j} \; = \; \frac{1}{F_2} \, \frac{1}{2\pi i}
\int^{c + i\infty}_{c - i\infty} d\nu \tilde{H}(\nu) \tilde{{\cal F}}_2(x/x_j,
{\textstyle \frac{1}{4}} - \nu,Q^2) \tilde{f}(x_j, {\textstyle \frac{1}{4}} +
\nu) .
\label{bzb}
\end{equation}
The Mellin transform $\tilde{f}(z,\omega)$ satisfies the transformed BFKL
equation (\ref{be}) which has the solution
\begin{equation}
\tilde{f}(z,\omega) \; = \; (k^2_0)^{-\omega} \bar{f}_0(\omega)
z^{-\tilde{K}(\omega)} ,
\label{bzc}
\end{equation}
see Eq.\ (\ref{bk}).
Similarly the solution of the equation for $\tilde{{\cal F}}_2$ is
\begin{equation}
\tilde{{\cal F}}_2(z,\omega,Q^2) \; = \; (Q^2)^{1-\omega} \bar{{\cal
F}}^{(0)}_2(\omega) z^{-\tilde{K}(\omega)} ,
\label{bzd}
\end{equation}
recall Eq.\ (\ref{bl}).  The explicit form of the
function $\bar{{\cal F}}_2^{(0)}(\omega)$
can be simply deduced from, for example, Appendix B of Ref.\ \cite{KMS}.
Inserting solutions Eq.\ (\ref{bzc}) and Eq.\ (\ref{bzd})
into Eq.\ (\ref{bzb}), we obtain
\begin{equation}
x_j \frac{\partial E_T}{\partial x_j} \; = \; \frac{1}{F_2} \; \frac{1}{2\pi i}
\int^{c + i\infty}_{c - i\infty} d\nu \,
\tilde{H}(\nu) \frac{(Q^2)^{\frac{3}{4}
+ \nu}}{(k^2_0)^{\frac{1}{4} + \nu}} \bar{{\cal F}}_2^{(0)}({\textstyle
\frac{1}{4}} - \nu) \bar{f}_0({\textstyle \frac{1}{4}} + \nu) \left(
\frac{x_j}{x} \right)^{\tilde{K}(\frac{1}{4} - \nu)} \left( \frac{1}{x_j}
\right)^{\tilde{K}(\frac{1}{4} + \nu)} .
\label{bze}
\end{equation}
We see that in the limit where $x_j/x \gg 1$ and $1/x_j \gg 1$, and in
particular in the central region away from the current jet and the proton
remnants where $x_j \sim x^{\frac{1}{2}}$ (or to be more precise where
$ \log (1/x_j) \sim \log (1/x^{\frac{1}{2}}))$, that the integral is dominated
by contributions from the region with $\nu \sim 0$.  We therefore expand the
relevant quantities about $\nu = 0$ and evaluate the resulting Gaussian
integral.  We find the master analytic formula
\begin{equation}
x_j \frac{\partial E_T}{\partial x_j} \; = \; C(Q^2\hat{k}^2)^{\frac{1}{4}}
x^{-\hat{\epsilon}} \exp \left( - \frac{ \{ |\hat{\lambda}^{\prime}|
\log (x^2_j/x) + \hat{T} \}^2}{2\hat{\lambda}^{\prime\prime}\log (1/x)}
\right) ,
\label{bzf}
\end{equation}
which may be compared with result
(\ref{br}) obtained by the simplified analytic
treatment.  Here $\hat{T} = \log (Q^2/\hat{k}^2)$.  We see we have
recovered  the same general structure as before, that is a Gaussian $E_T$
distribution as a  function of $ \log x_j$, with the characteristics (i) to
(iv)
listed in Sec.\ \ref{sec:twoa}.
The quantities are now more precisely determined.
First the parameter  $\hat{\epsilon}$, which controls the growth of the
distribution with decreasing  $x$, is given by
\begin{equation}
\hat{\epsilon} \; = \; \tilde{K}({\textstyle \frac{1}{4}}) -
\tilde{K}({\textstyle \frac{1}{2}}) \; = \; 2\bar{\alpha}_s \log  2 ,
\label{bzg}
\end{equation}
which would reduce to the previous result, $\epsilon =
\lambda^{\prime\prime}/32$, if the quadratic form Eq.\ (\ref{bj})
of $\tilde{K}(\omega)$
were valid to $\omega = \frac{1}{4}$.  The parameters $\hat{\lambda}^{\prime}$
and $\hat{\lambda}^{\prime\prime}$ denote the derivatives of $\tilde{K}$ at
$\omega = \frac{1}{4}$
\begin{equation}
\hat{\lambda}^{\prime} \; \equiv \; \tilde{K}^{\prime}({\textstyle
\frac{1}{4}}), \hspace*{1cm} \hat{\lambda}^{\prime\prime} \; \equiv \;
\tilde{K}^{\prime\prime}({\textstyle \frac{1}{4}}) ,
\label{bzh}
\end{equation}
and finally the normalization
\begin{equation}
C \; = \; \tilde{H}(0) \frac{(\lambda^{\prime\prime})^{\frac{1}{2}} \bar{{\cal
F}}^{(0)}_2(\frac{1}{4})\bar{f}_0(\frac{1}{4})}
{(\hat{\lambda}^{\prime\prime})^{
\frac{1}{2}} \bar{{\cal F}}^{(0)}_2(\frac{1}{2})\bar{f}_0(\frac{1}{2})}
\label{bzi}
\end{equation}
where
\begin{equation}
\tilde H(0) \; = \;
\bar{\alpha}_s {\Gamma^4({1 \over 4}) \over 2 \pi \Gamma^2({1 \over 2})}
\; \approx \; 8.75 \bar{\alpha}_s .
\label{bzia}
\end{equation}

As before we can translate the Gaussian $E_T$
distribution in $ \log x_j$ to one in
terms of rapidity $y$.  Again the distribution turns out to have the form of
Eq.\ (\ref{bzf}),
\begin{equation}
\frac{\partial E_T}{\partial y} \; = \; C(Q^2 \hat{k}^2)^{\frac{1}{4}}
x^{-\hat{\epsilon}} \exp \left( \frac{- \{ |\hat{\lambda}^{\prime}|2y +
\hat{T} + \frac{1}{2} \hat{\lambda}^{\prime 2} \log (1/x) -
\frac{1}{2}|\hat{\lambda}^{\prime}|\hat{T} \}
^2}{2\hat{\lambda}^{\prime\prime}\log (1/x)} \right) ,
\label{bzj}
\end{equation}
but we see that there are additional, non-leading, terms in the exponential.

\section{$E_T$ flow with strong $k_T$ ordering imposed}
\label{sec:three}

We compare the BFKL forms of the $E_T$ distribution, Eq.\
(\ref{br}) and Eq.\ (\ref{bzf}), with that  obtained from
conventional dynamics based on Altarelli-Parisi evolution in the
small $x$ limit.   To make a strict comparison we take fixed
$\alpha_s$.  Thus  rather than inserting into Eq.\ (\ref{bn}) the
BFKL solutions (\ref{bh}) and  (\ref{bi}), we replace  them by
the double leading logarithm (DLL) forms
\begin{equation}
zg(z,k^2) \; \sim \; \exp \left(2 \left[ \bar{\alpha}_s \log  \left(
\frac{k^2}{Q^2_0} \right) \log  \frac{1}{z} \right]^{\frac{1}{2}} \right)
\label{ca}
\end{equation}

\begin{equation}
{\cal F}(z,k^2,Q^2) \; \sim \; k^2 \, \exp \left(2 \left[  \bar{\alpha}_s
\log  \left( \frac{Q^2}{k^2} \right) \log  \frac{1}{z}
\right]^{\frac{1}{2}} \right) .
\label{cb}
\end{equation}
We substitute these
forms into Eq.\ (\ref{bn})
and Eq.\ (\ref{bm}), and change the $k^2_j$ integration
variable to $t = \log (k^2_j/Q^2_0)$.  We then obtain
\begin{equation}
x_j \frac{\partial E_T}{\partial x_j} \; \sim \; \frac{\bar{\alpha}_s}{F_2}
(Q^2_0)^{\frac{1}{2}} \int^T_0 dt \, \exp \left( \frac{t}{2} + 2\left[
\bar{\alpha}_s(T-t)\log  \left( \frac{x_j}{x} \right) \right]^{\frac{1}{2}}
+ 2 \left[ \bar{\alpha}_s t \log  \frac{1}{x_j} \right]^{\frac{1}{2}}
\right)
\label{cc}
\end{equation}
where $T \equiv \log (Q^2/Q^2_0)$.

It is informative to evaluate the $E_T$ distribution for $x_j = \sqrt{x}$, that
is for
\begin{equation}
\log  \frac{1}{x^2_j} \; = \; \log  \frac{1}{x} \; \equiv \; Y ;
\label{cd}
\end{equation}
a point \lq\lq midway" between the current jet and the proton remnants.  This
choice facilitates an analytic study and allows us to gain insight into the $x$
and $Q^2$ dependence of the $E_T$ flow.  When $ \log (x_j/x) = \log  (1/x_j)$
we can evaluate Eq.\ (\ref{cc}) using saddle point methods.  We find
\begin{equation}
\left .
x_j \frac{\partial E_T}{\partial x_j}\right |_{x_j = \sqrt{x}} \; \sim  \;
\bar{\alpha}_s(Q^2_0)^{\frac{1}{2}} \exp \left( {\textstyle \frac{1}{2}}
\beta T + \sqrt{2\bar{\alpha}_s TY} \left[ \sqrt{\beta} + \sqrt{1-\beta} -
\sqrt{2} \right] \right)
\label{ce}
\end{equation}
where $\beta$ is a known function of the ratio $T/2\bar{\alpha}_sY$.  To be
precise
\begin{equation}
\beta \; = \; {\textstyle \frac{1}{2}}
\left( 1 + \left\{ 1-4 \left[ 1 + \sqrt{1
+ T/2\bar{\alpha}_sY} \right] ^{-2}\right\} ^{\frac{1}{2}} \right) .
\label{cf}
\end{equation}
The $-\sqrt{2}$ in the square brackets
in Eq.\ (\ref{ce}) arises from substituting
\begin{equation}
F_2(x,Q^2) \; \sim \; \exp \left( 2 \left[ \bar{\alpha}_s \log
\left(\frac{Q^2}{Q^2_0} \right) \log  \left( \frac{1}{x} \right)
\right]^{\frac{1}{2}} \right)
\label{cg}
\end{equation}
into Eq.\ (\ref{cc}), rather than using Eq.\ (\ref{bt}).
After some algebra, it is
possible to show that the approximate estimate of the integral in
Eq.\ (\ref{cc})
predicts a slow increase of $E_T$ with decreasing $x_j$.

We inspect the properties of the $E_T$ flow, Eq.\ (\ref{ce}),
in two different limits.
First in the double leading logarithm limit, where $\alpha_sTY \gg 1$ but
$\alpha_sT \sim 1$ and $\alpha_sY \sim 1$, we see that $\beta \rightarrow 1$,
and so
\begin{equation}
\left .
x_j \frac{\partial E_T}{\partial x_j} \right |_{x_j = \sqrt{x}} \; \sim \;
(Q^2)^{\frac{1}{2}} \, \exp \left( -(\sqrt{2}-1) \sqrt{2\bar{\alpha}_2 TY}
\right) .
\label{ch}
\end{equation}
Thus in the DLL limit we find that the $E_T$ distribution grows essentially as
$(Q^2)^{\frac{1}{2}}$ with increasing $Q^2$.  In the second limit, $1/x
\rightarrow \infty$ at finite $Q^2$, we see that $\beta \rightarrow
\frac{1}{2}$.  Thus we find
\begin{equation}
\left .
x_j \frac{\partial E_T}{\partial x_j}
\right |_{x_j = \sqrt{x}} \; \sim \; e^{T/4}
\; \sim \; (Q^2Q^2_0)^{\frac{1}{4}} .
\label{ci}
\end{equation}
Moreover it is possible to show from Eq.\ (\ref{ce})
that with decreasing $x$ the distribution
decreases towards its final limit, which is indicated by Eq.\ (\ref{ci})
(modulo slowly varying logarithmic factors).
This should be contrasted with the BFKL behaviour
of Eq.\ (\ref{bzf})
which increases as $x^{-\hat{\epsilon}}$ with decreasing $x$.

\section {Calculation of $E_T$ flow}
\label{sec:four}

Above we have obtained analytic forms for the leading behaviour of the $E_T$
distribution in the $x_j \rightarrow 0$, $x/x_j \rightarrow 0$ limit, for fixed
$\alpha_s$.  These expose characteristic features of the expected $E_T$ flow in
deep-inelastic scattering
in the central region between, but well away from, the
current jet and proton remnants.  To obtain a more realistic estimate we solve
the appropriate BFKL equations numerically, and use running $\alpha_s$.  From
Eqs\ (\ref{ac}) and (\ref{bm})
we see that the $E_T$ flow, for a given $x,Q^2$, can be
expressed as
\begin{equation}
x_j \frac{\partial E_T}{\partial x_j} \; = \; \frac{1}{F_2} \int
\frac{dk^2_p}{k^2_p} \int \frac{dk^2_{\gamma}}{k^2_{\gamma}} \left(
\frac{3\alpha_s}{\pi} \right) {\cal F}_2 \left( \frac{x}{x_j},k^2_{\gamma},Q^2
\right) f(x_j,k^2_p) \int \frac{d\phi/2\pi}{\sqrt{k^2_p + k^2_{\gamma} +
2k_pk_{\gamma}\cos \phi}} .
\label{da}
\end{equation}
In this paper we focus on the central region with $x_j < 10^{-2}$
and $x/x_j <  10^{-1}$.  Thus to calculate the $E_T$ flow we have
to solve three different  BFKL equations, one for $f$ which
resums the lower gluon ladders of Fig. 2(b), one for ${\cal F}_2$
which resums the upper gluon ladders, and finally one for
$F_2(x,Q^2)$.  To be precise we determine $f(x_j,k^2_p)$ for
$x_j<10^{-2}$ by solving the BFKL equation, as described in
Ref.\cite{KMS}, starting from a gluon distribution at
$x_j=10^{-2}$ obtained from the MRS set of partons\cite{MRSD}.
The quantity ${\cal  F}_2/k^2_{\gamma}$ may be identified with
the structure function of a gluon of (approximate) virtuality
$k_{\gamma}^2$, integrated over its longitudinal momentum.  The
function ${\cal F}_2(z,k^2_{\gamma},Q^2)$ is determined for
$z<10^{-1}$ by solving the BFKL equation, as described in
Ref.\cite{KMS2}, starting from the boundary condition  ${\cal
F}_2={\cal F}_2^{(0)}$ at $z=10^{-1}$, where ${\cal F}_2^{(0)}$
is the contribution resulting just from the quark box (and
crossed box) in Fig. 2(b).  Finally $F_2(x,Q^2)$ is  determined
by the convolution of the gluon ladder with the quark box (and
crossed box), which symbolically may be written $F_2 = f \otimes
{\cal  F}_2^{(0)}$, as described in \cite{AKMS}.

The infrared cut-off on the transverse momenta integrations is
taken to be  $k^2_0 = 1$ GeV$^2$ throughout, a value which makes
the calculated values of  $F_2$ consistent with the recent
observations at HERA \cite{H1,ZEUS}.  The  predictions for the
$E_T$ flow are less sensitive to the choice of the cut-off  than
those of $F_2$.  The ultraviolet cut-off on the integrations over
the  transverse momenta is taken to be $10^4$ GeV$^2$; the
results for $f, {\cal  F}_2$ and $F_2$ are not sensitive to
reasonable variations about this value.  We  shall see below,
however, that the $E_T$-weighted integration of Eq.\ (\ref{da})
is  sensitive to the ultraviolet cut-off as $x_j$ decreases
towards the current jet.

Numerical results for the $E_T$ flow are presented in Fig.\ 3 for
$Q^2 = 10$  GeV$^2$ and values of $x = 10^{-8}, 10^{-7},...
10^{-4}$.  We show results for  such low values of $x$ so as to
establish a sufficiently large central region in  order to
compare the general features of the distribution with the
expectations  of the leading order analytic treatment of Sec.\
\ref{sec:two} From Fig.\ 3  we indeed  see a broad  Gaussian
shape in $ \log x_j$, with a peak below $x_j = \sqrt{x}$ and a
width which grows with decreasing $x$.  Moreover the full
numerical treatment  shows an {\it increase} of the height of the
distribution with decreasing $x$,  but not as rapid as
anticipated by the leading order analytic result.

We also show in Fig.\ 3 the $E_T$ distributions obtained using
Altarelli-Parisi  (AP) evolution for two values of $x$.  These
confirm the expectation that they  {\it decrease} with decreasing
$x$.  We come back to the BFKL comparison with AP  when
discussing Fig.\ 5, but first in Fig.\ 4 we explore the
sensitivity to the  choice of the ultraviolet cut-off in Eq.\
(\ref{da}).

Having established the general features of the $E_T$ flow, we
focus on the  distribution for $x = 10^{-4}$, a value which is
accessible at HERA.  In Fig.\ 4  the sequence of dashed curves
show the effect of choosing the ultraviolet  cut-off in Eq.\
(\ref{da})  to be $k^2_{\text{uv}} = 10^2, 10^3,... 10^7$
GeV$^2$, and the  continuous curve is the result obtained if the
cut-off is chosen to be $Q^2/z$  as implied by energy-momentum
conservation \cite{FHS}, where $z = x/x_j$.  At  relatively low
values of $Q^2$, such as $Q^2 = 10$ GeV$^2$, which are relevant
to experiments at HERA, energy conservation has a significant
effect on the  $E_T$ flow arising from BFKL gluon emission.
Fig.\ 5 compares the $E_T$ flow  for $Q^2 = 10$ GeV$^2$ with that
for $Q^2 = 100$ GeV$^2$ with, in each case,  $k^2_{\text{uv}} =
Q^2/z$.  Also shown are the expectations from Altarelli-Parisi
evolution.  As anticipated we see a significant difference
between the $E_T$  flow arising from BFKL and AP gluon emission,
particularly at lower $Q^2$  values.

We may use Eq.\ (\ref{bu}) to translate the $E_T$ flow as a
function $x_j$ into a  distribution in terms of the rapidity $y$
in the virtual-photon-proton c.m.\  frame.  For a reasonably flat
$E_T$ distribution a good estimate of $y$ is  obtained if we
insert the average value of $k^2_j$ into Eq.\ (\ref{bu}).   For
$Q^2 = 10$ GeV$^2$  we see that $E_T \approx 2$ GeV per unit of
rapidity and so we take  $k^2_j = 4$ GeV$^2$.  The result for $y$
is the scale shown above Fig.\ 5.  If  the $E_T$ distribution
observed by the H1 collaboration \cite{FEL} is compared  with
Monte Carlo expectations, then reasonable agreement is found for
$x >  10^{-3}$, but for $x < 10^{-3}$ more $E_T$ appears to be
present in the region  between the current jet and the proton
remnants than is implied by the favoured  Monte Carlo
calculations (which at present do not allow for BFKL emissions).
The deficiency is most evident in the rapidity region $x_j
\gtrsim 10^{-2}$,  which extends beyond the central region shown
in Fig.\ 5.  However it is  encouraging that a simple
extrapolation of the BFKL curves of Fig.\ 5 into this  region
indicates a sizeable enhancement  of the predicted $E_T$
distribution over  that expected from conventional AP dynamics.

\section{Conclusions}
\label{sec:five}

Our main objective is to find characteristics which are unique to
BFKL dynamics  at small $x$.  One striking feature of BFKL
dynamics is the very strong  $x^{-\lambda}$, with $\lambda
\approx 0.5$, increase of the deep-inelastic  structure function
$F_2(x,Q^2)$ as $x$ decreases in the HERA regime
\cite{AKMS,AKMS1}.  However this increase can be mimicked by
conventional  Altarelli-Parisi evolution from a very low starting
scale $Q^2_0 = 0.3$ GeV$^2$  \cite{GRV}.  Global or inclusive
quantities like structure functions do not  provide conclusive
tests of BFKL dynamics and it has been advocated that the  BFKL
effect can, in principle, be more definitively identified by
studies of the  final state in deep-inelastic scattering.  One
possibility is Mueller's proposal  \cite{MUE} of measuring the
$x/x_j$ dependence of deep-inelastic events  containing a
forward-going identified jet of longitudinal momentum $x_jp$.
Another, which we have studied here, is to measure the $E_T$ flow
accompanying  deep-inelastic events at small $x$.  A distinctive
feature of BFKL dynamics is  the diffusion pattern of the
transverse momentum flow, in contrast to the  strong-ordering of
$k_T$ towards $Q^2$ which is characteristic of conventional
Altarelli-Parisi dynamics.  Indeed it is the absence of strong
$k_T$-ordering  which leads to a strongly enhanced $E_T$
distribution at small $x$.  We analysed  this effect both
analytically and numerically, focussing attention on the  central
region between the current jet and the proton remnants.

In Sec.\ \ref{sec:three}  we performed a detailed analytic
analysis of the $E_T$ flow.  We  found that BFKL dynamics with
fixed $\alpha_s$ gives a broad Gaussian $E_T$  distribution as a
function of rapidity (or rather $ \log x_j$) which grows as
$x^{-\epsilon}$ with decreasing $x$ where $\epsilon =
(3\alpha_s/\pi)$2log2.   This should be contrasted with the much
smaller $E_T$ flow obtained assuming  strong $k_T$-ordering,
which gives an $E_T$ distribution that decreases with  decreasing
$x$, for fixed $Q^2$.  In Section 4 we confirmed these
qualitative  features using numerical solutions of the
appropriate BFKL equations.  As  expected we  found that the
$E_T$ flow which accompanies deep-inelastic events is  much
larger for BFKL than for conventional AP dynamics.  We also
showed, at the  low $Q^2$ values relevant to HERA, that energy
conservation should be used to  limit the energy-weighted
integrations over transverse momentum.

Taking, for example, deep-inelastic events with $x = 10^{-4}$ and
$Q^2 = 10$  GeV$^2$ we found that the $E_T$ flow in the central
rapidity region $(x_j <  10^{-2}, \, x/x_j < 10^{-1})$  between
the current jet and the proton remnants is  about 2 GeV per unit
of rapidity.  This result is very encouraging for the
identification of BFKL dynamics.  However for a quantitative
comparison with  experiment we will need to include contributions
to the $E_T$ flow coming from  the hadronization of the produced
partons, not only from the BFKL ladder, but  also from the
partons radiated from the current jet.  Also there is a
contribution radiated from the colour string which connects the
current jet to  the proton remnants.   An analysis of this
problem is under study and its results  will be reported
elsewhere.

\acknowledgements

We thank Yuri Dokshitzer and Genya Levin for useful discussions.  JK and KGB
thank Grey College and the Department of
Physics at the University of Durham for
warm hospitality.  Two of us (JK, ADM) thank the European Community for
Fellowships and two of us (KGB, PJS) thank the Polish
KBN - British Council collaborative research programme for partial support.
This work has also been supported in part by Polish KBN grant
no. 2 0198 91 01,  by the UK Science and
Engineering Research Council and EU under contract no.\ CHRX-CT92-0004.



\begin{figure}
\caption{
Deep-inelastic scattering at small $x$, accompanied by
soft gluon emission with total transverse energy
$\Sigma E_{Ti}$.}
\label{fig:one}
\end{figure}

\begin{figure}
\caption{
(a) Diagrammatic representation of Eq.\ (\ref{ac});
(b) explicit display of the gluon ladders which
are resummed by the BFKL equations for $f$ and
${\cal F}_i$.}
\label{fig:two}
\end{figure}

\begin{figure}
\caption{
The $E_T$ flow as a function of $\log x_j$, calculated using BFKL forms
in Eq.\ (\ref{da}),
for $x = 10^{-8},10^{-7},...10^{-4}$ and $Q^2 = 10$ GeV$^2$, in the central
region $(x_j < 10^{-2}, \, x/x_j < 10^{-1})$ between the current jet $(x_j
\approx x)$ and the proton remnants $(x_j \approx 1)$.  For comparison, the
dashed curves show the $E_T$ flow calculated using Altarelli-Parisi evolution
for $x = 10^{-6}$ and $10^{-4}$. }
\label{fig:three}
\end{figure}

\begin{figure}
\caption{
The sensitivity of the $E_T$ flow (for $x = 10^{-4}$ and $Q^2 = 10$ GeV$^2$) to
the choice of the ultraviolet cut-off $k^2_{\text{uv}}$ imposed on the
integrations over the
transverse momenta in Eq.\ (\ref{da}).  The dashed curves correspond
to $k^2_{\text{uv}} = 10^7,10^6,...10^2$ GeV$^2$, and the continuous curve to
$k^2_{\text{uv}} = Q^2/z$ with $z = x/x_j$. }
\label{fig:four}
\end{figure}

\begin{figure}
\caption{
The $E_T$ flow for $x = 10^{-4}$
calculated using BFKL forms (continuous curves)
and conventional AP dynamics (dashed curves) for two different values of $Q^2$,
namely $Q^2 = 10$ and 100 GeV$^2$.
The BFKL predictions are obtained from Eq.\ (\ref{da})
with $k^2_{\text{uv}} = Q^2/z$ with $z = x/x_j$.  The variable $\log x_j$ is
related to the virtual-photon-proton c.m.\ rapidity, $y$, via Eq.\ (\ref{bu}).
The
approximate rapidity scale for $Q^2 = 10$ GeV$^2$ is shown above the figure.
}
\label{fig:five}
\end{figure}
\end{document}